\documentclass[preprint]{aastex}

\begin{document}
\title{Oxygen Overabundance in the Extremely Iron-Poor Star CS~29498--043\footnote{Based on data collected
at the Subaru Telescope, which is operated by the National Astronomical
Observatory of Japan.}}

\author{Wako Aoki\altaffilmark{2}, John E. Norris\altaffilmark{3},
Sean G. Ryan\altaffilmark{4}, Timothy C. Beers\altaffilmark{5}, \\
Norbert Christlieb\altaffilmark{6}, Stelios Tsangarides\altaffilmark{4},
Hiroyasu Ando\altaffilmark{2}}

\altaffiltext{2}{National Astronomical Observatory, Mitaka, Tokyo,
181-8588 Japan; email: aoki.wako@nao.ac.jp, ando@optik.mtk.nao.ac.jp}
\altaffiltext{3}{Research School of Astronomy and Astrophysics, The
Australian National University, Mount Stromlo Observatory, Cotter
Road, Weston, ACT 2611, Australia; email: jen@mso.anu.edu.au}
\altaffiltext{4}{Department of Physics and Astronomy, The Open
University, Walton Hall, Milton Keynes, MK7 6AA, UK; email:
s.g.ryan@open.ac.uk, s.tsangarides@open.ac.uk} 
\altaffiltext{5}{Department of Physics and Astronomy, Michigan State
University, East Lansing, MI 48824-1116; email: beers@pa.msu.edu}
\altaffiltext{6}{Hamburger Sternwarte, Universit\"{a}t Hamburg,
Gojenbergsweg 112, 21029 Hamburg, Germany; nchristlieb@hs.uni-hamburg.de}

\begin{abstract} 

An abundance analysis for the carbon-enhanced, extremely iron-poor ([Fe/H]$\sim
-3.5$) star CS~29498--043 has been obtained using new high-resolution, high
signal-to-noise spectra from the Subaru Telescope. The [\ion{O}{1}] forbidden
line at 6300~{\AA} and the \ion{O}{1} triplet feature at 7771-7776~{\AA} are
both clearly detected. The overabundance of oxygen is significant ([O/Fe]$>2$).
In addition, Na, Co, and Ni abundances have been newly measured. The abundance
pattern from C to Ni of this object is quite similar to that of CS~22949--037,
another extremely metal-poor star with large excesses of C, N, O, and the
$\alpha$-elements. The abundance patterns of these two stars suggest the
existence of supernovae progenitors that ejected relatively little material from
their iron cores during the very early era of nucleosynthesis in the Galaxy. The
metallicity in these objects, when one includes the elements C, N, and O in the
tally of total metals, is not as low as in the most metal-poor stars, suggesting
the existence of quite different formation processes for these iron-deficient objects than
pertain to the bulk of other metal-deficient stars.

\end{abstract} 

\section{Introduction}\label{sec:intro}

The surface chemical composition of extremely metal-poor stars is
believed to reflect the yields of heavy elements produced by the first
generations of massive stars in our Galaxy. Reported large variations
of the elemental abundance patterns of some species (in particular
those associated with carbon and the neutron-capture elements) found
in very metal-poor stars (e.g., McWilliam et al. 1995; Ryan, Norris, \& Beers
1996; Cayrel et al. 2004) suggest a diversity of the
nucleosynthesis and explosion mechanisms of the supernovae which were
presumably responsible.

Our previous study of the extremely metal-poor star CS~29498--043
([Fe/H]$\sim-3.7$)\footnote{[A/B] = $\log(N_{\rm A}/N_{\rm B})-
\log(N_{\rm A}/N_{\rm B})_{\odot}$, and $\log \epsilon_{\rm A} =
\log(N_{\rm A}/N_{\rm H})+12$ for elements A and B.} revealed
remarkable overabundances of C, N, Mg, and Si in this object
\citep{aoki02a,aoki02b}. The chemical nature of this object, which is
similar to that of another extremely metal-poor star, CS~22949--037 (McWilliam
et al. 1995, Norris, Ryan \& Beers 2001; Depagne et al. 2002), suggests
that these objects formed from the yields of supernovae that ejected
relatively little material from the regions surrounding their iron
cores at the time of their explosion \citep[e.g.,
][]{tsujimoto03,umeda03a}. A remaining important question is the
origin of the large overabundances of C and N, because these two
elements are expected to be significantly affected during the
evolution of low-mass stars. In order to distinguish the contributions
of low- and high-mass stars, determinations of the O abundance, as
well as those of the $\alpha$ elements, are quite important. The O
abundance is also vital to estimate the metallicity of this object. In
this paper, we report a new analysis of the chemical composition
of CS~29498--043, including the O abundance, derived from
high-resolution, high signal-to-noise spectra obtained with the Subaru
Telescope.

\section{Observations and measurements}\label{sec:obs}

The observations reported herein were obtained with the Subaru
Telescope High Dispersion Spectrograph \citep[HDS; ][]{noguchi02},
with a spectral resolving power $R = 50,000$, in October 2002 and May
2003. The wavelength coverage and signal-to-noise ratios per pixel
(S/N) are listed in Table~\ref{tab:obs}. The EEV-CCD with 13.5~$\mu$m
pixels was used with the $2\times$2 binning mode.  Standard data
reduction procedures were carried out with the IRAF echelle
package\footnote{IRAF is distributed by the National Optical Astronomy
Observatories, which is operated by the Association of Universities
for Research in Astronomy, Inc. under cooperative agreement with the
National Science Foundation.}.

For the abundance analysis described herein, we combined these spectra
with a similar-quality blue spectrum obtained in our previous study
\citep{aoki02a}.  Equivalent widths of absorption lines for 12 metals
(14 ionization species) were measured by fitting Gaussian profiles,
and are listed in Table \ref{tab:ew} along with the line data used in the
analysis.


In order to investigate the binarity of this object, which might have relevance
for the interpretation of the origin of the abundance peculiarities, we measured
heliocentric radial velocities ($V_{\rm r}$) for each spectrum, as given in
Table~\ref{tab:obs}. The measurements were made using clean, isolated
\ion{Fe}{1} lines. Our measurements for spectra obtained in previous observing
runs are also given in Table~\ref{tab:obs}. No clear variation of the radial
velocities is found for this object, although the $V_{\rm r}$ values are
slightly decreasing over the past three years. Further monitoring of radial
velocity to determine if it is in fact a long-period binary is clearly
desirable.

\section{Analysis and Results}\label{sec:ana}

An LTE abundance analysis was carried out using the model atmospheres of
\citet{kurucz93}. An estimate of the effective temperature, $T_{\rm
eff} = 4570K$, was obtained from the $V-K$ color index, using the
\citet{alonso99} empirical temperature scale, adopting the
photometry data from \citet{norris99} ($V=13.72$) and from the 2
Micron All Sky Survey Point Source Catalog \citep{skrutskie97} ($K
=10.94$), along with an interstellar reddening of $E(B-V)=0.09$,
estimated from the \citet{schlegel98} map. We adopted $T_{\rm
eff}=4600$~K for the following abundance analysis.


After the analysis was completed, $R$ and $I$ photometry data also
became available, obtained with the ESO-Danish 1.54\,m telescope and
DFOSC. The derived colors, $V-R=0.615$ and $V-I=1.222$ (Cousins
system), result in temperature estimates of $T_{\rm eff}=4570$~K and
4500~K, respectively, if Alonso et al. (1999)'s scale is applied.
These results support our effective temperature determined from the
$V-K$ color.

In our previous analysis \citep{aoki02a,aoki02b}, we adopted $T_{\rm
eff}=4400$~K as a preferred value, taking account of the correlation
between the derived Fe abundances and the lower excitation potential
of \ion{Fe}{1} lines found in the result for $T_{\rm
eff}=4600$~K. However, this correlation is not clear in the present
analysis, which includes new \ion{Fe}{1} lines detected in the revised
spectrum covering both blue and red ranges, especially the lines with
high excitation potentials ($\sim 4.3$~eV). For this reason, we have
adopted $T_{\rm eff}=4600$~K as the preferred value, but consider 4400~K
as an alternative choice.

Using the model atmospheres of \citet{kurucz93} for the adopted
effective temperatures, we performed abundance analyses in the
standard manner for the measured equivalent widths.  Surface gravities
($\log g$) were determined from the ionization balance between Fe
{\small I} and Fe {\small II}; the microturbulence ($v_{\rm tur}$) was
determined from the Fe {\small I} lines by demanding no dependence of
the derived abundance on equivalent widths. The effect of large
enhancements of C, N, Mg, and Si were included in the calculation of
chemical equilibrium for the abundance analysis. The excess of oxygen
found in the present analysis, as noted below, was also taken into
account. We note that these effects are not included in the
calculations of the thermal structure of atmospheres, but they are
estimated to be small for such low metallicity stars. The final derived
atmospheric parameters are $\log g=1.2, v_{\rm tur}=2.4$~km~s$^{-1}$ and
[Fe/H]$=-3.5$.

The derived abundances ([X/Fe] and [Fe/H]) are listed in
Table~\ref{tab:res}.  The sixth column provides the results when
$T_{\rm eff}=4400$~K is adopted. The surface gravity, determined by
the same methods, but for $T_{\rm eff}=4400$~K, is $\log g=0.6$. The
same microturbulence and metallicity as for $T_{\rm eff}=4600$~K were
used. For comparison, in the last column of this table, we also provide
the chemical abundances derived for the atmospheric parameters adopted
by \citet{israelian03,israelian04} ($T_{\rm eff}=4300$, $\log g=1.5$ and
[Fe/H]$=-3.5$). Since the iron abundances from \ion{Fe}{1} and
\ion{Fe}{2} derived by our (LTE) analysis show a significant
discrepancy (0.5~dex), we assumed [Fe/H]$=-3.5$, which is similar to the
value derived from \ion{Fe}{2} lines, to derive [X/Fe] values for
individual elements.

\subsection{Oxygen}

\begin{figure} 
\includegraphics[width=8.5cm]{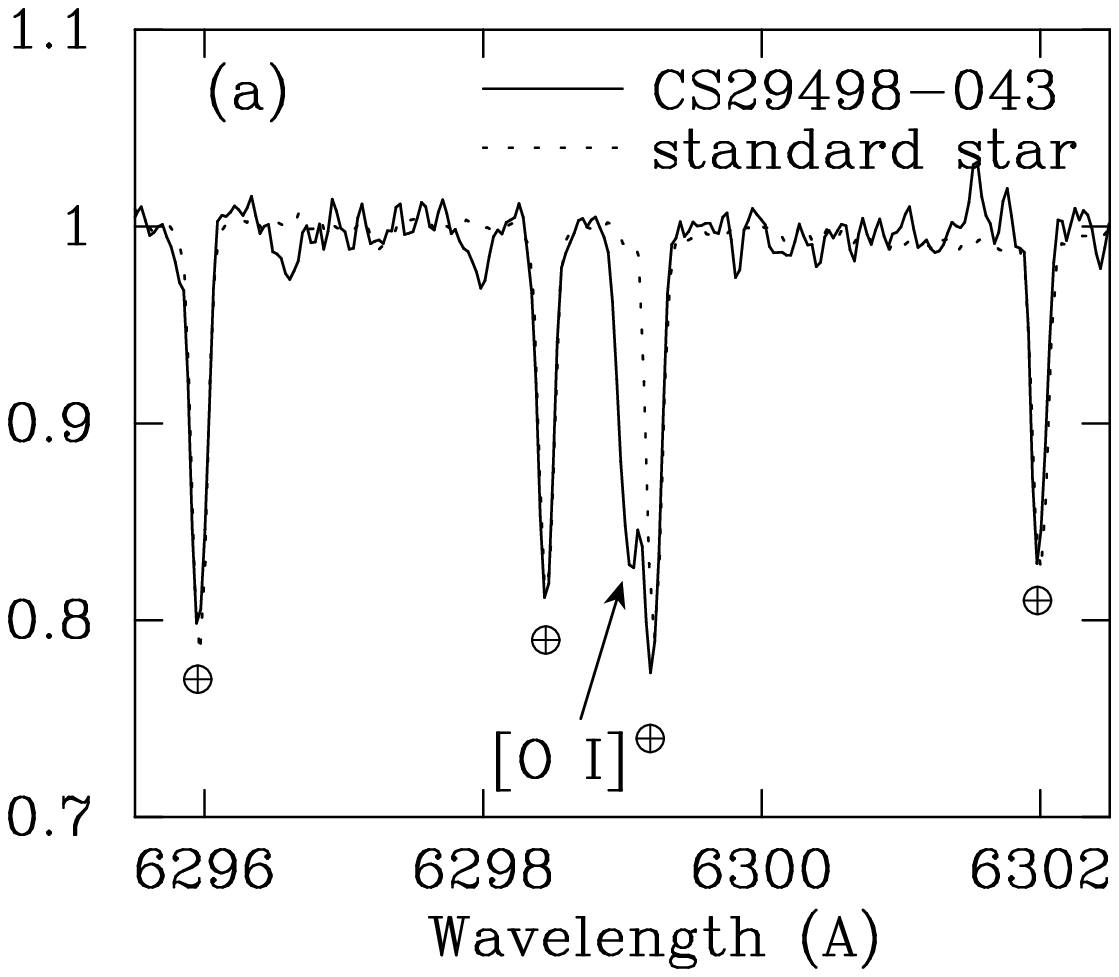} 
\includegraphics[width=8.5cm]{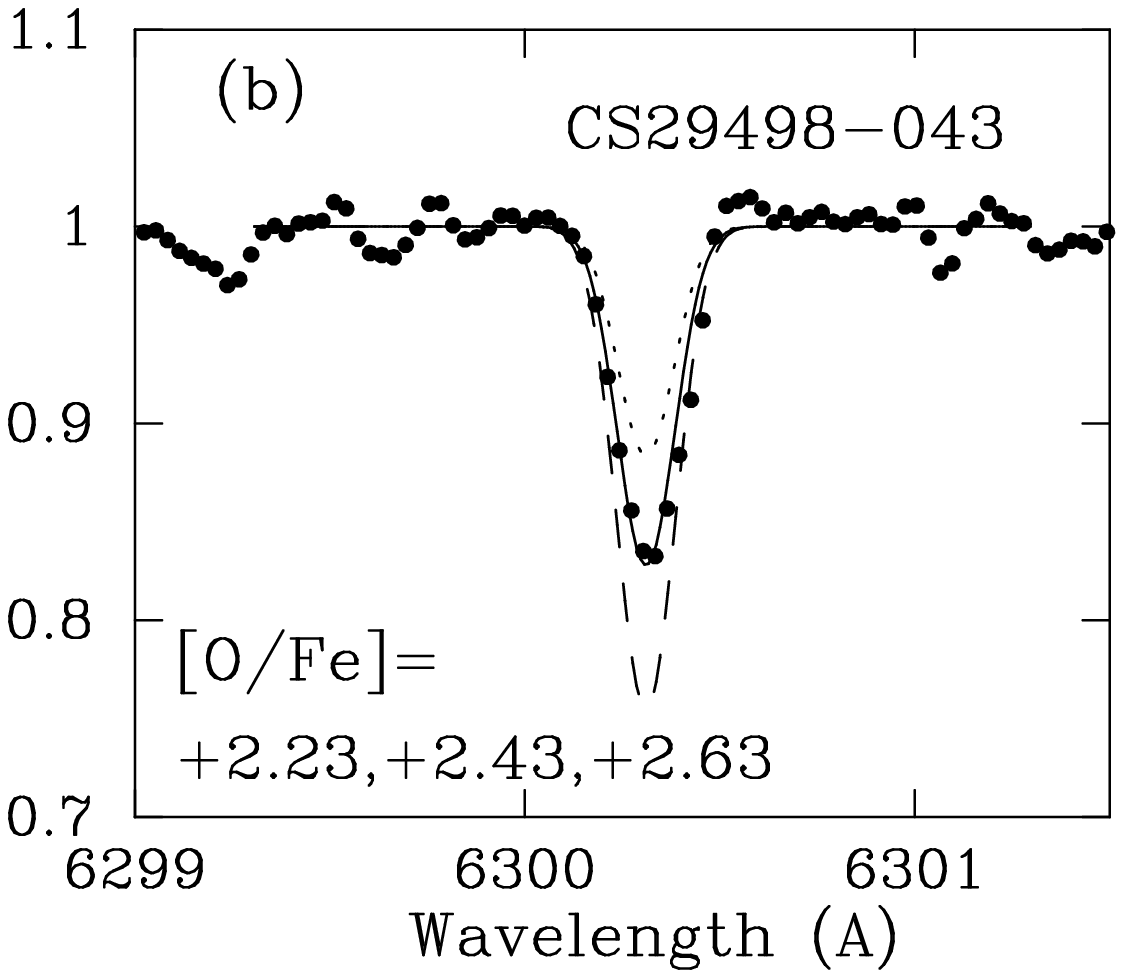}
\caption[]{{\it a}: A comparison of the spectra of CS~29498--043 and
HD~178840 (standard star). Doppler shifts in the spectra are not corrected for. The
telluric absorption lines are indicated by plus-circles. {\it b}: The
spectrum of CS~29498--043, corrected for the Doppler shift and telluric
lines (dots). Synthetic spectra assuming [O/Fe]$=2.23, 2.43$, and 2.63 are
shown by dotted, solid, and dashed lines, respectively.}
\label{fig:o1f} 
\end{figure}

The [\ion{O}{1}] 6300~{\AA} line is clearly detected in the spectrum of
CS~29498--043. Since, at this wavelength, the October 2002 spectrum is
significantly affected by strong [\ion{O}{1}] emission in the earth's
atmosphere, the analysis was made using the May 2003 spectrum, which provided a
more favorable velocity shift. Figure~\ref{fig:o1f}a shows a comparison of the
spectra of CS~29498--043 and a standard star (HD~178840) observed just after the
observation of CS~29498--043 with a similar zenith distance. We confirmed that
the telluric lines are corrected by the spectrum of the standard star within
1~\% of the continuum level for the lines at 6295.9, 6298.4, and 6302.0~{\AA}.
This indicates that telluric absorption does not significantly affect the
measurement of the equivalent width of the [\ion{O}{1}] line, which is estimated
to be 31.4 m{\AA} (Table 2). The spectrum of this star, corrected for the
telluric absorption lines (primarily H$_{2}$O) and the Doppler shift estimated from
clean \ion{Fe}{1} lines, is shown as the heavy dots in Figure~\ref{fig:o1f}b. The lines
are synthetic spectra for the three assumed oxygen abundances presented in the
figure\footnote{We adopt a solar oxygen abundance of $\log
\epsilon$(O)$_{\odot}$=8.74 determined from the 6300~{\AA} forbidden
line by \citet{nissen02}, which is consistent with
\citet{allendeprieto01} within the uncertainty.}.

The equivalent width of the [\ion{O}{1}] 6363~{\AA} line measured for
our spectrum is 14~m{\AA}, which provides a consistent oxygen
abundance with that obtained from the [\ion{O}{1}] 6300~{\AA}. However, the
quality of the observed spectrum at 6363~{\AA} is not as good as that
at 6300~{\AA}, and a few unidentified lines with similar strengths to
the [\ion{O}{1}] 6363~{\AA} line appear around this
wavelength. Therefore, we adopt the result obtained from the
[\ion{O}{1}] 6300~{\AA} line for the oxygen abundance derived from
forbidden lines.

\begin{figure} \includegraphics[width=12cm]{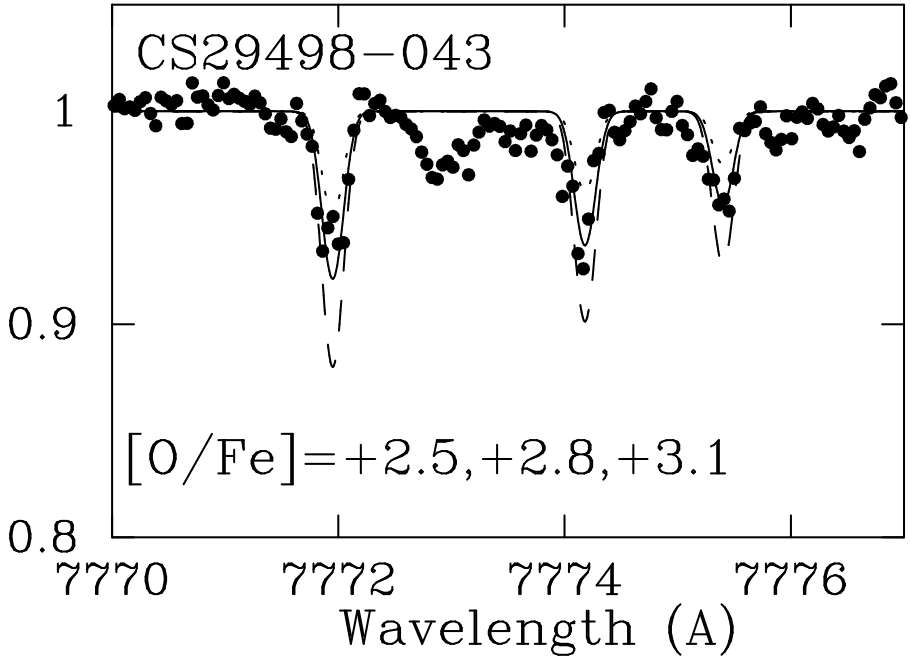}
\caption[]{The same as Figure 1b, but for the \ion{O}{1} triplet lines.}
\label{fig:o1ir} \end{figure}

The \ion{O}{1} triplet lines at 7770--7776~{\AA} are also clearly
detected in CS~29498--043. Figure \ref{fig:o1ir} shows the observed
and synthetic spectra for this wavelength range. We note that the
observed spectrum in this wavelength range is, unfortunately, affected
by fringes due to interference occurring at the surface of the CCD
\citep{noguchi02}. However, our equivalent widths agree well with
those measured by \citet{israelian04} for their Keck/HIRES
spectrum. The oxygen abundance derived from the triplet is 0.47~dex
higher than that obtained from the forbidden line
(Table~\ref{tab:res}). However, it should also be noted that the
triplet lines are known to be significantly affected by NLTE
effects. \citet{takeda03} estimated the NLTE correction for the oxygen
abundance derived from the triplet lines to be about $-0.1$~dex for
giants with $T_{\rm eff}=4500$~K. This correction reduces the
discrepancy between the oxygen abundances from the triplet and
forbidden lines to about 0.4~dex. Similar discrepancies have also been
found in other very metal-poor ([Fe/H]$<-2$) stars studied in detail
by \citet{takeda03}. Hence, we do not pursue the origin of this
problem further.  We note that the discrepancy is larger ($\sim 0.6$
dex, after NLTE correction) if the lower effective temperature
(4400~K) is adopted, since the strengths of the high excitation
triplet lines are quite sensitive to temperature. We adopt the oxygen
abundance derived from the forbidden line in the following discussion,
because the result is insensitive to the temperature and NLTE
corrections.

\citet{israelian03} recently reported a large O overabundance
for CS~29498-043 from an analysis of the \ion{O}{1} triplet ([O/Fe]=+3.12),
adopting $T_{\rm eff}$=4300~K. The oxygen abundance derived by our LTE analysis
for the triplet lines adopting their atmospheric parameters is [O/Fe]=+3.36.
They reported the NLTE correction for the oxygen abundance derived from the
triplet lines to $-0.18$~dex for this object; after applying this correction,
our results agree very well with theirs. 

In addition, \citet{israelian04} also reported the oxygen abundance
from the forbidden line that is about 0.5~dex lower than that from the
triplet line (the discrepancy is about 0.8~dex after the NLTE
correction is applied). This large discrepancy, compared with that
found in our above analysis for $T_{\rm eff}=4600$~K (0.4~dex), is
likely due to the low assumed effective temperature
(4300~K). \citet{israelian03,israelian04} have also detected the \ion{O}{1}
triplet in the extremely metal-poor star CS~22949--037, which also
exhibits similar overabundances in C, N, O, and the $\alpha$-elements
as in CS~29498--043. They reported an even more significant
discrepancy (1.2~dex) between the oxygen abundances derived from the
triplet and the forbidden line \citep{depagne02} for
CS~22949--037. Given the large discrepancy of oxygen abundances in
this object, and also in CS~29498--043, further detailed discussion of
O abundances in these objects is difficult. However, it remains clear
that the O overabundances are remarkable in these two stars.

\subsection{Other Elements}

A standard analysis using model atmospheres was performed for the measured
equivalent widths for most of the other detectable elements, while a spectrum
synthesis technique was applied for the C$_{2}$ and CN molecular bands, and the
Al and Mn lines which exhibit contamination from CH molecular lines. The
analysis technique and molecular line data are the same as reported in
\citet{aoki02a,aoki02b}. An exception is that the effect of hyperfine
splitting is included in the analysis of Mn and Co lines, using the
line list of \citet{mcwilliam95} and \citet{pickering96},
respectively, in the present work.

The results are listed in Table~\ref{tab:res}. The abundances of the
species Co and Ni, which were not detected in the previous spectrum,
are reported here for the first time.  We searched for MgH $A-X$ lines in the
region 5100--5200~{\AA}, but found no clear feature of this
molecule. Since the strength of the strongest MgH lines expected for
the Mg abundance in Table~\ref{tab:res} is about 1~m{\AA}, spectra
with higher quality are required to measure the MgH features.

Random abundance errors in the analysis are estimated from the standard
deviation of the abundances derived from individual lines for each species.
These values are sometimes unrealistically small, however, when only a few lines
are detected. For this reason, we adopted the larger of (a) the value for the
listed species and (b) that for \ion{Fe}{1} (0.09~dex) as estimates of the
random errors. Errors arising from uncertainties of the atmospheric parameters
were evaluated using $\sigma (T_{\rm eff})=200$~K, $\sigma (\log g)=0.5$, and
$\sigma (v_{\rm tur}) =0.5$~km s$^{-1}$. Finally, we derived the total
uncertainty by adding in quadrature the individual errors, and list them in
Table \ref{tab:res}.

NLTE effects on our abundance analysis are estimated from the study of
metal-poor giants by \citet{gratton99}. They derived the abundance corrections
([X/H]$_{\rm NLTE}-$[X/H]$_{\rm LTE}$) for Fe (derived from \ion{Fe}{1} lines),
Na (from the D-lines) and Mg to be typically +0.1, +0.0, and +0.2~dex,
respectively, for stars with $T_{\rm eff}=$4000--5000~K and $\log g=1.5$. The
correction for Fe abundance propagates to a change of estimated surface gravity,
which is determined from the ionization equilibrium between \ion{Fe}{1} and
\ion{Fe}{2}, by about 0.3~dex. Though this correction will make a
systematic change of the abundance results, the effect is still within
the uncertainties estimated above.

\section{Discussion and concluding remarks}

\begin{figure} \includegraphics[width=12cm]{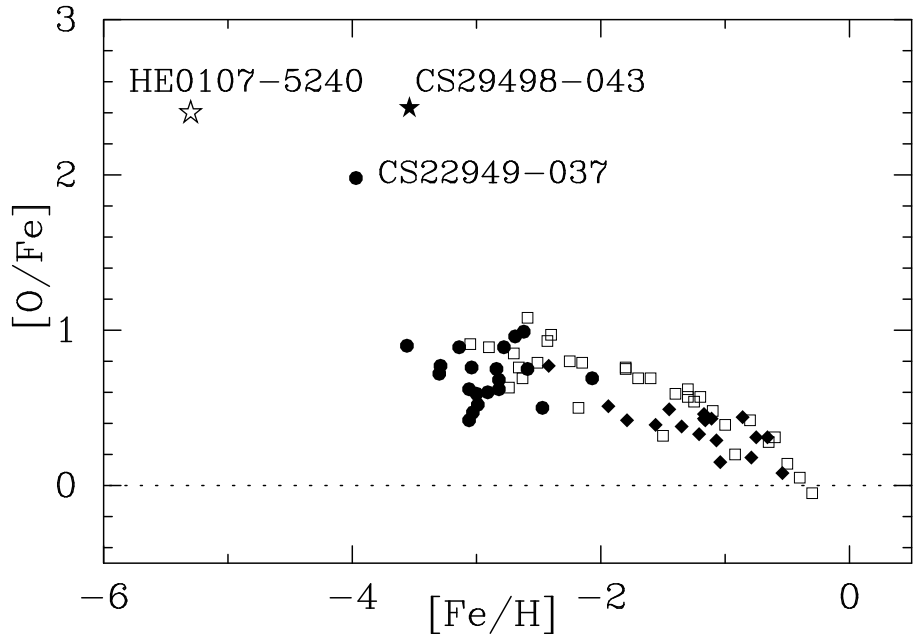} \caption[]{[O/Fe]
as a function of [Fe/H] for CS~29498--043 (the filled star, this work)
and others in the literature: open squares: \citet{israelian01}; filled
diamonds: \citet{nissen02}; filles circles: \citet{cayrel03}; the open
star: \citet{bessell04}. The filled symbols show the [O/Fe] values
from the [\ion{O}{1}] line, while open ones mean those from triplet
lines or OH molecular lines.}  \label{fig:ofe} \end{figure}

Recent studies of O abundances in very metal-poor stars, within the
framework of one-dimensional model atmospheres, have revealed the
presence of a possible increase of O/Fe ratios with decreasing
metallicity in the range of [Fe/H]$<-2$.  Figure \ref{fig:ofe} shows
[O/Fe] as a function of [Fe/H] for our object and for other stars
described in the literature \citep{israelian01, nissen02, cayrel03,
bessell04}. The filled symbols show the [O/Fe] values determined from
the [\ion{O}{1}] line, while open symbols indicate those obtained from
either the triplet lines or OH molecular lines. A discrepancy between
the results from the different indicators is seen, as discussed in
subsection 3.1. However, the O overabundance of CS~29498--043, as well
as that of CS~22949--037 \citep{depagne02}, is unusually high,
compared with the other very metal-poor stars ([O/Fe]$\lesssim
+1.0$). The [O/Fe] of the most iron-deficient star presently known,
HE~0107--5240 ([Fe/H] = -5.3; Christlieb et al. 2002, 2004) is also as
high as found in these two stars \citep{bessell04}. We note that, even
though three of the four objects with [Fe/H]$<-3.5$ have
[O/Fe]$\gtrsim +2.0$, there are stars in this metallicity range in
which oxygen lines are {\it not} detected, and hence whose oxygen
abundances are not as high as in these three stars. Though it remains
possible that [O/Fe] values continuously increase with decreasing
[Fe/H] for stars with [Fe/H]$<-3$, the unusually high oxygen
abundances in the above three stars are more likely to arise from a
different nucleosynthesis history from that pertaining to other
metal-deficient stars.

\begin{figure} \includegraphics[width=12cm]{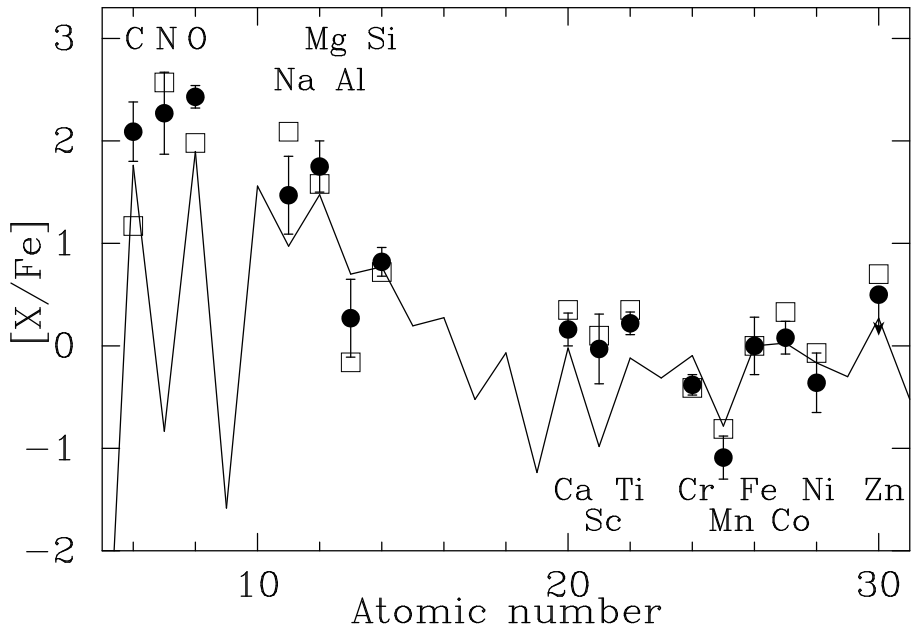}
\caption[]{Elemental abundance patterns, as a function of atomic
number, of CS~29498--043 (filled circles, this work) and CS~22949--037
(open squares, Depagne et al. 2002).  The solid line shows an
abundance pattern predicted from the Umeda \& Nomoto's supernova model
for a 30~$M_{\odot}$ star with [Fe/H]$=-4.0$ and the explosion energy
of $5\times 10^{52}$~erg, assuming the mixing region of $M_{\rm
r}=2.44$--12.6$M_{\odot}$ and the matter ejection factor of $f=0.004$
(Umeda \& Nomoto 2003b).}  \label{fig:res} \end{figure}

Figure \ref{fig:res} shows the relative elemental abundances ([X/Fe]),
as a function of atomic number, for CS~29498--043 and
CS~22949--037. The similarity of the abundance patterns between these
two stars is remarkable. The present analysis for the new spectrum of
CS~29498--043 confirms the similarity for O and Na, which show large
overabundances in both stars. The [O/Fe] value of CS~29498--043 is
higher, by 0.36~dex, than that of CS~22949--037, if the same solar O
abundance is adopted. However, we do yet not insist on a real
difference of the O abundances between the two stars, taking account
of the uncertainty in the O abundance determination (as seen, e.g., in
the discrepancy in the results between the [\ion{O}{1}] line(s) and
the O triplet). Recall that for the present analysis, the O abundances
in both objects were determined from the [\ion{O}{1}] 6300~{\AA} line.

The solid line in Figure \ref{fig:res} shows the abundance ratios
predicted by the supernova model of \citet{umeda03a} for a
30~$M_{\odot}$ star, assuming substantial mixing and fall-back (Umeda
\& Nomoto 2004). The large overabundances of C, O, Mg, and Si
relative to iron-peak elements are well reproduced by this model. The
large overabundance of N in these stars could be associated with the
operation of the CN-cycle during their evolution on the red-giant
branch \citep{depagne02}. Note as well that the Co/Fe and Ni/Fe
ratios of CS~29498--043 might be slightly lower than those of
CS~22949--037. The upper-limit on the Zn abundance of CS~29498--043 is
also lower than the Zn abundance of the other object. In the Umeda \&
Nomoto (2004) models, lower Co and Zn abundance ratios indicate a
smaller explosion energy and/or more effective mixing. More accurate
estimates of the abundance ratios of these elements in the present
objects, as well as for other similar stars that might be found in the
future, are desirable to better constrain the parameters of the
present explosion models.

Finally, we would like to consider the metallicity of these objects
and their extremely low iron abundances. We here define the
metallicity by the {\it total abundance} of C, N, O, Mg, Si, and Fe,
i.e., [A/H]=[(C+N+O+Mg+Si+Fe)/H]. The metallicities of CS~29498--043
and CS~22949--037, using this definition, are [A/H]$=-1.26$ and
$-2.00$, respectively. Oxygen is the most important contributor to the
metallicity in these stars (60--70\% of the metallic species). When
we adopt the above definition of metallicity, these three objects might
be regarded as {\it extremely iron-deficient stars} rather than {\it
extremely metal-poor stars}.

The metallicity of the most iron-poor ([Fe/H]$=-5.3$) star
HE~0107--5240 \citep{christlieb02, christlieb04} is [A/H]$=-1.77$,
using this definition, adopting [O/Fe]$=+2.4$ \citep{bessell04}, and
assuming [Si/Fe]=0.0. In this case, however, the dominant contributor
to the metallicity is carbon (about 95\%). The abundance pattern of
this object can also be explained by supernova models by
\citet{umeda03a}, assuming a larger mixing region and a smaller matter
ejection factor. Although this object shows no clear overabundance of
$\alpha$-elements, the origin of its peculiar abundance pattern may be
related to those of CS~29498--043 and CS~22049--037.

The extremely iron-deficient ([Fe/H]$<-3.5$) stars found in previous
studies (Norris et al. 2001, Carretta et al. 2002, Franc\c{o}is et
al. 2003), in general, {\it do not exhibit} large excesses of lighter
elements, except for CS~22949--037 (Norris et al. 2001). These stars
can be regarded as extreme cases of metal-deficient stars with higher
iron abundance ($-3.5<$[Fe/H]$\lesssim-2.5$), which would form from
interstellar matter polluted by first-generation supernovae. Their
extremely low metallicity suggests high explosion energies of the
supernovae that provided metals, and induced the formation of
second-generation stars \citep[e.g., ][]{cioffi88}.  Hence, the
formation mechanism of the above three iron-deficient stars with
excesses of C and O may be quite different from the other
iron-deficient stars.  Even though their total atmospheric
metallicities are rather high ([A/H]$\gtrsim -2$), these three stars
are also expected to be produced from material polluted by
first-generation supernovae, because of their peculiar abundance
patterns.  However, the extremely low iron abundances of these objects
might be regarded as resulting from the particular yields of their
progenitors, rather than from mixing processes within the interstellar
matter from which they were formed. The relatively high total
metallicity of these objects may indicate low explosion energies of
the supernovae which induced the formation of these objects.

If the species C and O were present in the interstellar matter of the
birth clouds of these objects they would supply important cooling
sources, in particular for HE~0107--5240, hence their overabundances
may help understand the possibility of low-mass star formation in the
early Galaxy, as suggested by \citet{umeda03a} (see also Bromm \& Loeb
2003). This may also explain the existence of the plethora of C-rich
objects amongst extremely iron-deficient stars (Norris, Ryan, \& Beers
1997; Beers 1999; Rossi et al. 1999). Further abundance studies of
stars with extremely low iron abundances are indispensable to
establish the proper classification of these stars, and to understand
the formation mechanism(s) of early-generation stars in our Galaxy.

\acknowledgments


We are grateful to F.-J. Zickgraf for carrying out the photometric
observations ($R$- and $I$-Band), and to B.  Marsteller for the speedy
data reduction, which provided confirmatory information for our
temperature estimate of CS~29498-043. JEN acknowledges support from
grant DP0342613 of the Australian Research Council. TCB acknowledges
partial support from grants AST 00-98508 and AST 00-98549 awarded by
the US National Science Foundation. NC is grateful to the American
Astronomical Society for financial support through a Chretien
International Research Grant.

\clearpage

\begin{deluxetable}{lcccc}
\tablewidth{0pt}
\tablecaption{SUBARU OBSERVATIONS FOR CS~29498--043\label{tab:obs}}
\startdata
\tableline
\tableline
Obs. Date (JD)  & Wavelength & Exposure  & S/N\tablenotemark{a} & Radial velocity  \\
                &   ({\AA})  &   (min)   &                      & (km s$^{-1}$)    \\
\tableline
Sep. 2000 (2,451,802)\tablenotemark{b} & 3770--4760 & 270  & 45  & $-32.50\pm 0.18$  \\
July 2001 (2,452,115)\tablenotemark{c} & 3550--5250 & 120  & 87  & $-32.60\pm 0.46$  \\
Oct. 2002 (2,452,573)                  & 5100--7800 & 80   & 112  & $-32.74\pm 0.42$  \\
May  2003 (2,452,785)                  & 5100--7800 & 90 & 148  & $-32.95\pm 0.39$  \\
May  2003 (2,452,786)                  & 3550--5250 & 90 & 77   & $-33.10\pm 0.45$  \\
\tableline
\enddata

\tablenotetext{a}{ S/N ratios per 1.8~km~s$^{-1}$ pixel at 4320~{\AA}
and 6270~{\AA} for the blue and red spectra, respectively.  }

\tablenotetext{b}{Based on spectra obtained with AAT/UCLES}

\tablenotetext{c}{
Results of this observing run have already been reported in Aoki et al. (2002a,b). 
}
\end{deluxetable}

\begin{deluxetable}{lccc}
\tablecaption{EQUIVALENT WIDTHS \label{tab:ew}}
\startdata
\tableline
\tableline
$\lambda$({\AA}) & $\log gf$ & $\chi$(eV) & $W$(m{\AA})\\
\tableline
  O I    &            &            &          \\
 6300.30 &     -9.820 &      0.000 &    31.4s \\
 6363.78 &    -10.303 &      0.020 &    14.0x \\
 7771.95 &      0.358 &      9.147 &    19.5s \\
 7774.18 &      0.212 &      9.147 &    15.6s \\
 7775.40 &     -0.010 &      9.147 &    10.9s \\
  Na I   &            &            &          \\  
 5889.95 &      0.101 &      0.000 &    204.5 \\
 5895.92 &     -0.197 &      0.000 &    180.8 \\
  Mg I   &            &            &          \\  
 4057.50 &     -0.890 &      4.346 &     65.0 \\
 4571.10 &     -5.688 &      0.000 &    109.9 \\
 4702.99 &     -0.520 &      4.346 &    101.1 \\
 5172.69 &     -0.380 &      2.712 &    232.9x \\
 5183.60 &     -0.160 &      2.717 &    253.4x \\
 5528.40 &     -0.490 &      4.346 &    102.8 \\
  Al I   &            &            &          \\  
 3961.52 &     -0.340 &      0.010 &    136.5s \\
  Si I   &            &            &          \\  
 4102.94 &     -2.910 &      1.909 &     75.5 \\
  Ca I   &            &            &          \\  
 5588.76 &      0.358 &      2.526 &     12.5 \\
 6102.72 &     -0.770 &      1.879 &      5.9 \\
 6122.22 &     -0.320 &      1.886 &     21.4 \\
  Sc II  &            &            &          \\  
 5526.79 &      0.020 &      1.768 &      6.0 \\
  Ti I   &            &            &          \\  
 4991.07 &      0.436 &      0.836 &     17.5 \\
 5173.74 &     -1.062 &      0.000 &     10.0 \\
 5210.38 &     -0.828 &      0.048 &     12.4 \\
  Ti II  &            &            &          \\  
 4468.52 &     -0.600 &      1.131 &     90.9 \\
 4501.27 &     -0.760 &      1.116 &     73.9 \\
 4533.97 &     -0.770 &      1.237 &     92.4 \\
 4563.77 &     -0.960 &      1.221 &     64.4 \\
 4571.96 &     -0.530 &      1.572 &     87.1 \\
 5185.90 &     -1.350 &      1.893 &      8.3 \\
 5188.69 &     -1.210 &      1.582 &     30.0 \\
 5226.53 &     -1.300 &      1.566 &     26.5 \\
 5381.01 &     -1.970 &      1.566 &      5.6 \\
  Cr I   &            &            &          \\  
 5208.44 &      0.158 &      0.941 &     54.2 \\
 5409.79 &     -0.720 &      1.030 &      8.9 \\
  Mn I   &            &            &          \\  
 4030.75 &     -0.470 &      0.000 &     83.7s \\
 4033.07 &     -0.620 &      0.000 &     68.3s \\
  Fe I   &            &            &          \\  
 4005.24 &     -0.610 &      1.557 &     92.5 \\
 4063.59 &      0.060 &      1.557 &    116.9 \\
 4071.74 &     -0.022 &      1.608 &    112.4 \\
 4461.65 &     -3.210 &      0.087 &     86.3 \\
 4466.55 &     -0.600 &      2.832 &     33.1 \\
 4489.74 &     -3.966 &      0.121 &     40.8 \\
 4494.56 &     -1.136 &      2.198 &     36.8 \\
 4528.61 &     -0.822 &      2.176 &     59.3 \\
 4531.15 &     -2.155 &      1.485 &     35.3 \\
 4592.65 &     -2.449 &      1.557 &     14.9 \\
 4602.94 &     -2.210 &      1.485 &     37.0 \\
 4871.32 &     -0.360 &      2.865 &     42.0 \\
 4890.75 &     -0.390 &      2.876 &     30.5 \\
 4891.49 &     -0.110 &      2.851 &     48.0 \\
 4903.31 &     -0.930 &      2.882 &     10.0 \\
 4918.99 &     -0.340 &      2.865 &     33.9 \\
 4920.50 &      0.070 &      2.833 &     56.0 \\
 4994.13 &     -2.956 &      0.915 &     26.6 \\
 5006.12 &     -0.610 &      2.833 &     23.1 \\
 5012.07 &     -2.642 &      0.859 &     61.5 \\
 5041.07 &     -3.090 &      0.958 &     28.6 \\
 5041.76 &     -2.200 &      1.485 &     34.3 \\
 5049.82 &     -1.344 &      2.279 &     26.3 \\
 5051.63 &     -2.795 &      0.915 &     46.3 \\
 5068.77 &     -1.040 &      2.940 &     11.5 \\
 5083.34 &     -2.958 &      0.958 &     38.6 \\
 5110.41 &     -3.760 &      0.000 &     61.5 \\
 5123.72 &     -3.068 &      1.011 &     31.7 \\
 5127.36 &     -3.307 &      0.915 &     22.4 \\
 5151.91 &     -3.322 &      1.011 &     24.8 \\
 5166.28 &     -4.195 &      0.000 &     38.5 \\
 5171.60 &     -1.793 &      1.485 &     55.1 \\
 5191.46 &     -0.550 &      3.038 &     15.3 \\
 5192.34 &     -0.420 &      2.998 &     22.5 \\
 5194.94 &     -2.090 &      1.557 &     36.6 \\
 5202.34 &     -1.838 &      2.176 &     14.2 \\
 5216.27 &     -2.150 &      1.608 &     30.8 \\
 5232.94 &     -0.060 &      2.940 &     43.0 \\
 5254.96 &     -4.764 &      0.110 &     10.7 \\
 5266.56 &     -0.390 &      2.998 &     24.6 \\
 5269.54 &     -1.321 &      0.859 &    129.7 \\
 5328.04 &     -1.466 &      0.915 &    122.2 \\
 5328.53 &     -1.850 &      1.557 &     52.7 \\
 5332.90 &     -2.780 &      1.557 &     10.7 \\
 5341.02 &     -1.950 &      1.608 &     38.9 \\
 5369.96 &      0.540 &      4.371 &      9.0 \\
 5371.49 &     -1.645 &      0.958 &    111.5 \\
 5383.37 &      0.640 &      4.312 &      9.7 \\
 5397.13 &     -1.993 &      0.915 &     99.6 \\
 5405.77 &     -1.844 &      0.990 &     99.9 \\
 5415.20 &      0.640 &      4.386 &      8.9 \\
 5424.07 &      0.520 &      4.320 &     14.6 \\
 5429.70 &     -1.879 &      0.958 &    101.0 \\
 5434.52 &     -2.122 &      1.011 &     80.8 \\
 5446.92 &     -1.910 &      0.990 &     95.2 \\
 5455.61 &     -2.098 &      1.011 &     92.9 \\
 5497.52 &     -2.849 &      1.011 &     39.3 \\
 5506.78 &     -2.797 &      0.990 &     43.3 \\
 5572.84 &     -0.275 &      3.397 &     17.2 \\
 5586.75 &     -0.096 &      3.368 &     21.0 \\
 5615.64 &      0.050 &      3.332 &     24.4 \\
 6065.48 &     -1.530 &      2.609 &     11.4 \\
 6136.61 &     -1.400 &      2.453 &     18.9 \\
 6230.72 &     -1.281 &      2.559 &     17.1 \\
 6393.60 &     -1.432 &      2.433 &     20.2 \\
  Fe II  &            &            &          \\  
 4508.28 &     -2.580 &      2.856 &     11.8 \\
 4520.23 &     -2.600 &      2.807 &      6.2 \\
 4583.83 &     -2.020 &      2.807 &     30.1 \\
 4923.93 &     -1.320 &      2.891 &     51.2 \\
 5018.45 &     -1.220 &      2.891 &     62.5 \\
 5234.62 &     -2.270 &      3.221 &      7.9 \\
  Co I   &            &            &          \\  
 3894.07 &      0.100 &      1.049 &     49.8 \\
 3995.31 &     -0.220 &      0.923 &     36.0 \\
 4092.39 &     -0.940 &      0.923 &     19.5 \\
 4121.32 &     -0.320 &      0.923 &     50.4 \\
  Ni I   &            &            &          \\  
 3597.70 &     -1.115 &      0.212 &    100.6 \\
 3612.74 &     -1.423 &      0.275 &     70.8 \\
 3619.39 &      0.020 &      0.423 &    105.3 \\
 3775.57 &     -1.408 &      0.423 &     59.5 \\
 5476.91 &     -0.890 &      1.826 &     25.9 \\
  Zn I   &            &            &          \\
 4722.15 &     -0.390 &      4.030 &    $<7$  \\
 4810.53 &     -0.170 &      4.080 &    $<7$  \\
  Sr II  &            &            &          \\  
 4077.71 &      0.150 &      0.000 &    106.4 \\
  Ba II  &            &            &          \\  
 4554.03 &      0.163 &      0.000 &     85.4 \\
 4934.09 &     -0.160 &      0.000 &     76.2 \\
 5853.67 &     -1.020 &      0.604 &      5.9 \\
 6141.70 &     -0.070 &      0.704 &     30.5 \\
\tableline
\enddata
\tablenotetext{a}{The symbol 'x' indicates the lines measured, but excluded from the abundance analysis. The symbol 's' means the equivalent widths estimated from the spectrum synthesis.}
\end{deluxetable}

\begin{deluxetable}{ccccccc}
\tablewidth{0pt}
\tablecaption{ABUNDANCE RESULTS FOR CS~29498--043\label{tab:res}}
\startdata
\tableline
\tableline
Element \hspace{0.4cm} & [X/Fe] & $\log\epsilon_{\rm X}$ & $n$ &  $\sigma$ &  [X/Fe]$_{4400}$\tablenotemark{a} & [X/Fe]$_{\rm I03}$\tablenotemark{b} \\ 
\tableline
$^{12}$C (C$_{2}$)\dotfill & $+$2.09 & 7.10   & syn    & 0.29 & $+$2.30 & $+$2.65 \\
N (CN)         \dotfill    & $+$2.27 & 6.70   & syn    & 0.40 & $+$2.48 & $+$2.43  \\
O ([O I]6300) \dotfill     & $+$2.43 & 7.63   & 1(syn) & 0.11 & $+$2.38 & $+$2.38 \\
O (triplet)   \dotfill     & $+$2.90 & 8.10   & 3(syn) & 0.34 & $+$3.11 & $+$3.36 \\
Na I          \dotfill     & $+$1.47 & 4.25   &  2     & 0.38 & $+$1.50 & $+$0.83 \\
Mg I          \dotfill     & $+$1.75 & 5.79   &  4     & 0.25 & $+$1.91 & $+$1.46 \\
Al I          \dotfill     & $+$0.27:& 3.20:  & 1(syn) & 0.38 & $+$0.32: & $-$0.15:\\
Si I          \dotfill     & $+$0.82 & 4.84   &  1     & 0.14 & $+$0.90 & $+$0.55 \\
Ca I          \dotfill     & $+$0.16 & 2.97   &  3     & 0.16 & $+$0.25 & $-$0.17 \\
Sc II         \dotfill     & $-$0.03 &$-$0.47 &  1     & 0.34 & $-$0.04 & $-$0.08 \\
Ti I          \dotfill     & $+$0.22 & 1.62   &  3     & 0.11 & $+$0.16 & $-$0.32 \\
Ti II         \dotfill     & $+$0.38 & 1.78   &  9     & 0.32 & $+$0.39 & $+$0.38 \\
Cr I          \dotfill     & $-$0.38 & 1.77   &  2     & 0.10 & $-$0.38 & $-$0.83 \\
Mn I          \dotfill     & $-$1.09 & 0.90   & 2(syn) & 0.21 & $-$0.56 & $-$1.05 \\
Fe I ([Fe/H]) \dotfill     & $-$3.54 & 3.96   & 65     & 0.28 & $-$3.75 & $-3.94$ \\
Fe II ([Fe/H]) \dotfill    & $-$3.53 & 3.97   &  6     & 0.34 & $-$3.71 & $-3.42$ \\
Co I          \dotfill     & $+$0.06 & 1.43   &  4     & 0.16 & $+$0.00 & $-$0.42 \\
Ni I          \dotfill     & $-$0.36 & 2.35   &  5     & 0.29 & $-$0.35 & $-$0.81 \\
Zn I          \dotfill     & $<+0.5$ & $<1.6$ &  2     &      & $<+$0.5 & $<+0.9$ \\
Sr II \dotfill             & $-$0.47 &$-$1.09 &  1     & 0.31 & $-$0.48 & $-$0.48 \\
Ba II \dotfill             & $-$0.46 &$-$1.78 &  4     & 0.20 & $-$0.52 & $-$0.58 \\
\tableline
\enddata

\tablenotetext{a}{Abundance results derived for $T_{\rm eff}=4400$K.}

\tablenotetext{b}{Abundance results derived for the stellar parameters adopted by Israelian et al. (2003; 2004)}

\end{deluxetable}

\clearpage

\begin{thebibliography}{}

\bibitem[Allende Prieto et al. (2001)]{allendeprieto01} Allende
Prieto, C., Lambert, D. L., \& Asplund, M., 2001, \apj, 556, L63

\bibitem[Alonso, Arribas, \& Mart\'{i}nez-Roger
(1999)]{alonso99}Alonso, A., Arribas, S., \& Mart\'{i}nez-Roger,
C. 1999, \aaps, 140, 261

\bibitem[Aoki et al. (2002a)]{aoki02a} Aoki, W., Norris, J.E., Ryan,
S.G., Beers, T.C., \& Ando, H. 2002a, \apj, 576, L141

\bibitem[Aoki et al. (2002b)]{aoki02b} Aoki, W., Norris, J.E., Ryan,
S.G., Beers, T.C., \& Ando, H. 2002b, \pasj, 54, 933

\bibitem[Beers et al. (1999)]{beers99} Beers, T.C. 1999, in Third
Stromlo Symposium: The Galactic Halo, eds. B. Gibson, T. Axelrod, \&
M. Putman (ASP: San Francisco), 165, p. 206

\bibitem[Bessell, Christlieb, \& Gustafsson (2004)]{bessell04}
Bessell, M. S., Christlieb, N., \& Gustafsson, B. 2004, submitted to
ApJ Letters

\bibitem[Bromm \& Loeb (2003)]{bromm03} Bromm, V., \& Loeb, A. 2003,
Nature, 425, 812

\bibitem[Carretta et al. (2002)]{carreta02} Carretta, E., Gratton, R.,
Cohen, J. G., Beers, T. C., \& Christlieb, N. 2002, \aj, 124, 481

\bibitem[Cayrel et al. (2004)]{cayrel03} Cayrel, R., Depagne, E.,
Spite, M., Hill, V., Spite, F., Francois, P., Plez, B., Beers, T.C.,
Primas, F., Andersen, J., Barbuy, B., Bonifacio, P., Molaro, P., \&
Nordstr\"{o}m, B. 2004, \aap, in press

\bibitem[Christlieb et al. (2002)]{christlieb02} Christlieb, N.,
Bessell, M.S., Beers, T.C., Gustafsson, B., Korn, A., Barklem, P.S.,
Karlsson, T, Mizuno-Wiedner, M., \& Rossi, S. 2002, Nature, 419, 904

\bibitem[Christlieb et al. (2004)]{christlieb04} Christlieb, N., 
Gustafsson, B., Korn, A.J., Barklem, P.S., Beers, T.C., Bessell, M.S.,
Karlsson, T., \& Mizuno-Wiedner, M. 2004, \apj, in press (astro-ph/0311173)

\bibitem[Cioffi, McKee \& Bertschinger (1988)]{cioffi88} Cioffi,
D. F., McKee, C. F., \& Bertschinger, E. 1988, \apj, 334, 252

\bibitem[Depagne et al. (2002)]{depagne02} Depagne, E., et al. 2002,
\aap, 390, 187

\bibitem[Franc\c{o}is et al. (2003)]{francois03}Franc\c{o}is, P. et
al. 2003, \aap, 403, 1105

\bibitem[Gratton et al. (1999)]{gratton99} Gratton, R. G., Carretta,
E., Eriksson, K. \& Gustafsson, B., 1999, \aap, 350, 955

\bibitem[Israelian et al. (2001)]{israelian01} Israelian, G., Rebolo,
R., Garc\'{i}a L\'{o}pez, R. J., Bonifacio, P., Molaro, P., Basri, G., \&
Shchukina, N. 2001, \apj, 551,833

\bibitem[Israelian et al. (2003)]{israelian03} Israelian, G.,
Shchukina, N., Rebolo, R., Basri, G., \& Gonz\'{a}lez Hern\'{a}nde\'{z},
J.I. 2003, in Carnegie Observatories Astrophysics Series, Vol. 4

\bibitem[Israelian et al. (2004)]{israelian04} Israelian, G.,
Shchukina, N., Rebolo, R., Basri, G., Gonz\'{a}lez Hern\'{a}nde\'{z},
J. I., Kajino, T., \& Nomoto, K.  2004, submitted to \aap

\bibitem[Kurucz (1993)]{kurucz93} Kurucz, R. L., 1993, CD-ROM 13,
ATLAS9 Stellar Atmospheres Programs and 2~km/s Grid (Cambridge:
Smithsonian Astrophys. Obs.)

\bibitem[McWilliam et al. (1995)]{mcwilliam95} McWilliam, A., Preston,
G. W., Sneden, C., \& Searle, L. 1995, \aj, 109, 2757

\bibitem[Nissen et al. (2002)]{nissen02} Nissen, P. E., Primas, F.,
Asplund, M., \& Lambert, D.L. 2002, \aap, 235, 251

\bibitem[Noguchi et al. (2002)]{noguchi02} Noguchi, K., Aoki, W.,
Kawanomoto, S., et al. 2002, \pasj, 54, 855

\bibitem[Norris, Ryan, \& Beers (1997)]{norris97} Norris, J.E.,
Ryan. S.G., \& Beers, T.C. 1997, \apj, 488, 350

\bibitem[Norris, Ryan, \& Beers (1999)]{norris99} Norris, J. E., Ryan,
S. G., \& Beers, T.  C. 1999, \apjs, 123, 639

\bibitem[Norris, Ryan, \& Beers (2001)]{norris01} Norris, J. E., Ryan,
S. G., \& Beers, T.  C. 2001, \apj, 561, 1034

\bibitem[Pickering (1996)]{pickering96} Pickering, J. C. 1996, \apjs,
107, 811

\bibitem[Rossi, Beers, \& Sneden (1999)]{rossi99} Rossi, S., Beers,
T.C., \& Sneden, C. 1999,in Third Stromlo Symposium: The Galactic
Halo, eds. B. Gibson, T. Axelrod, \& M. Putman (ASP: San Francisco),
165, p. 268

\bibitem[Ryan, Norris, \& Beers (1996)]{ryan96} Ryan, S.G., Norris,
J.E., \& Beers, T.C. 1996, \apj, 471, 254


\bibitem[Schlegel, Finkbeiner, \& Davis (1998)]{schlegel98} Schlegel,
D.J., Finkbeiner, D.P., \& Davis, M. 1998, \apj, 500, 525

\bibitem[Skrutskie et al. (1997)]{skrutskie97} Skrutskie, M.F., et
al. 1997, in The Impact of Large Scale Near-IR Sky Surveys,
ed. F. Garzon et al. (Dordrecht: Kluwer), p. 187

\bibitem[Takeda (2003)]{takeda03} Takeda, Y. 2003, \aap, 402, 343 

\bibitem[Tsujimoto \& Shigeyama (2003)]{tsujimoto03} Tsujimoto, T., \&
Shigeyama, T. 2003, \apj, 584, L87

\bibitem[Umeda \& Nomoto (2003)]{umeda03a} Umeda, H., \& Nomoto,
K. 2003, Nature, 422, 871

\bibitem[Umeda \& Nomoto (2004)]{umeda03b} Umeda, H., \& Nomoto,
K. 2004, \apj, submitted (astro-ph/0308029)

\end{thebibliography}
\end{document}